\documentstyle[12pt,epsfig]{article}
\def\rsim{\mathrel{\raise2pt\hbox to 8pt{\raise -5pt\hbox{$\sim$}\hss{$>$}}}}
\def\lsim{\mathrel{\raise2pt\hbox to 8pt{\raise -5pt\hbox{$\sim$}\hss{$<$}}}}
\begin{document}

\begin{center}
{\bf ASSESSMENT OF TRITON POTENTIAL ENERGY\\}
\vspace*{0.20in}
by\\
\vspace*{0.20in}
J.L.\ Friar\\
Theoretical Division\\
Los Alamos National Laboratory\\
Los Alamos, NM  87545, USA\\
\vspace*{0.20in}
and\\
\vspace*{0.20in}
G.L.\ Payne\\
Department of Physics and Astronomy\\
University of Iowa\\
Iowa City, IA  52242, USA\\
\vspace*{0.5in}
{\bf Abstract}
\end{center}

An assessment is made of the dominant features contributing to the triton
potential energy, with the objective of understanding qualitatively their
origins and sensitivities.  Relativistic effects, short-range repulsion, and
OPEP dominance are discussed.  A determination of the importance of various
regions of nucleon-nucleon separation is made numerically.

\pagebreak

Substantial and numerous recent successes[1] in the area of few-nucleon physics
indicate that it is appropriate to begin an assessment of our level of
understanding of the physics underlying the dynamics of few-nucleon systems. In
recent months and years, a variety of computational and theoretical techniques
have been successfully applied to the bound (or low-lying) states of A =
2,3,4,5, and 6 and to the scattering states of the trinucleon systems. 
Attempts are underway to extend this program and are likely to be successful.
These efforts have been very informative and generally produce results in
excellent agreement with experiment except for a few small, but potentially very
significant, discrepancies.  A vital question that must be asked is whether we
can now (or ever) understand the nuclear force well enough to accommodate all
that we have learned and hope to learn from future progress. A systematic
understanding of the structure of few-nucleon systems and unraveling the
dynamics of these systems have long been considered the {\it raison d'\^etre}
of the field of few-nucleon physics.

One recent calculation[2] underscores the recent and potential precision of the
field. The Nijmegen group have constructed several new potential models, whose
dominant feature is a charge-dependent one-pion-exchange potential (OPEP) [see
below], by directly fitting the model parameters to the nucleon-nucleon (NN)
scattering data. The partial-wave-dependent (Reid-like) versions of these
potentials fit the NN data so well that they can be considered as alternative
phase-shift analyses. Published and unpublished (preliminary) {\bf local}
versions of these potentials produce triton binding energies $\sim$ 7.62(3) MeV,
while nonlocal versions are more bound by $\sim$ 100 keV. The local result is
therefore a benchmark for gauging the size of small components of the nuclear
force. Since locality and nonlocality (which clearly exists and is conceptually
important) cannot be determined from data analysis, only theoretical arguments
can discriminate between them. Constructing a theoretically sound and
compelling nuclear force model is therefore a requirement before we can
discriminate between (small) three-nucleon forces and vagaries of the NN force
model. Thus we must address the question: ``Is this a realistic possibility?''

A major historical problem within the field, which has tainted the
aforementioned successes, is the old perception that the nuclear force is too
complicated to be understood in anything approaching fundamental terms and that
perturbation theory, which underlies virtually all theoretical approaches,
cannot be performed because the coupling constants are too large, implying that
any systematic development is impossible.  At least within the few-nucleon
systems, where ``exact'' or ``complete'' nonrelativistic solutions of the
Schr\"{o}dinger equation are now routine and dynamical details can be tested,
this pessimism is largely unfounded, which we argue below.

An appreciation of this problem begins with a dichotomy.  On the one hand the
exchange of a pion (OPEP) has long been the archetype of the nuclear force,
while on the other hand folklore (based entirely on properties of heavier
nuclei) attributes most nuclear binding to correlated two-pion exchange.  The
first indication that the latter property did not hold for few-nucleon systems
was given in Ref. [3], where it was shown that the tensor potential and the
$^3S_1 - ^3D_1$ partial wave of the nuclear force dominate the triton binding. 
Replacing the latter potential with a ``pure'' OPEP potential leads to nearly
the same triton binding energy, demonstrating that OPEP dominates triton
binding.  This was more clearly documented[2,4] by subsequent calculations of
$\langle V_{\pi} \rangle$ for the triton, alpha particle and other systems: 
the fraction of the potential energy from OPEP, $\langle V_{\pi} \rangle
/\langle V \rangle$, is 70-80\% for most realistic potentials. Not surprisingly,
OPEP has an even more dominant effect in the deuteron, where an adequate (but
not excellent) deuteron can be obtained using a pure regulated OPEP; ironically,
OPEP is too strong at short distances to make a good deuteron without the
regulation[3]. Details of the nuclear force are obviously required in order to 
guarantee excellent agreement with observables.

The importance of OPEP has been highlighted by the Nijmegen group[5], whose
phase-shift analysis of nucleon-nucleon scattering data included an explicit
OPEP tail for separations $r \geq$ 1.4 fm and a phenomenological treatment of
the shorter-range interaction.  Their study concluded that the NN data require
the exchanges of a charged pion with  mass $m_{\pi^{+}}\,c^2 =$ 139.4(10) MeV,
and a neutral pion with mass $m_{\pi^{0}}\,c^2 =$ 135.6(13) MeV.  Both results
are consistent with the free-particle masses, while the very small error bars
reflect the overall importance of OPEP.  Although it has been long known that
OPEP played a very significant role in some NN observables, this was the first
direct quantitative measure of its global importance.

Thus, it is not unfair to characterize OPEP as the ``Coulomb potential'' of
few-nucleon systems, although there remains a significant contribution from
shorter-range forces. This is exceedingly fortunate, because the former physics
is tractable, while the latter is exceedingly complex. These non-OPEP parts can
be further subdivided into two convenient classes:  two-pion-range (TPEP) and
short-range forces.  By two-pion exchange we mean the exchange of two
uncorrelated (noninteracting) pions, rather than strongly correlated resonances
such as the $\rho$. Chiral symmetry is expected to play a major role in such a
process, even in the long-range tail of the $\rho$-exchange channel. In
addition, isobars contribute to this part of the force, although many potential
models do not contain explicit isobar-induced components. Assessments of the
role of the long-range two-pion-exchange force have been very disappointing. 
The literature on the topic is replete with model calculations but few definite
conclusions.  Many potentials contain a TPEP in one form or another, but some
do not.  Theoretical arguments based on chiral symmetry suggest that it is weak
compared to OPEP[6,7].  An urgent need of the field is an assessment of what
(if any) long-range TPEP components are {\bf required} by the NN data, rather
than merely reflecting theoretical prejudice. Several recent developments
suggest that such a study may be tractable[5].

The inner part of the (non-OPEP) potential contributes 20-30\% of the triton
potential energy, and a detailed understanding of its origin may be problematic
unless the full power of QCD can be directly applied.  A variety of folklore
arguments have been constructed that suggest that an understanding is not
possible.  We address two of these arguments next: one concerning nonlocalities
and the other Z-graphs.

The only known methodology for constructing a (potentially) testable dynamical
framework in nuclear physics is field theory (from which OPEP follows).  Field
theory leads to powerful constraints on systems but, as conventionally
formulated, produces complications that seem far removed from our
understanding of nuclear dynamics.  Doubts are frequently expressed (although
seldom in print) about whether field theory is meaningful in the nuclear
context. Nevertheless, if further progress is to be made, it will almost 
certainly be within the field theory context.

Given that QCD is the theory of strong interactions, the most natural
description of any strongly interacting system would be in terms of the
``simple'' degrees of freedom (d.o.f.):  quarks and gluons.  Unfortunately, as
has been emphasized many times, a perturbation expansion of this theory in the
confinement region is not tractable[8,9].  Nucleons (and pions) are complicated
composites of these d.o.f. and have extended structure. Because their mutual
interactions are consequently nonlocal, it has been argued that field theory
cannot be meaningfully applied to nuclear physics. All of these statements 
are correct except the last one.

The second (related) argument concerns the way a field theory of nucleons is
organized, with manifest covariance playing a fundamental role. Nucleon and
antinucleon d.o.f.\ are combined and treated together. The presence of
nucleon-antinucleon pairs (Z-graphs, which have a very short range) has long
been a thorn in the community's side.  They play a huge and unphysical role in
some models of the nuclear force, and ``pair suppression'' was invoked long ago
to remove their effect in an {\it ad hoc} manner.  Subsequently, the argument
against pairs has been updated to accommodate QCD and modern particle
phenomenology.  The latter follows from the observation that $N \bar{N}$ (pair)
final states are not an important component in high-energy reactions, and
therefore pairs should not be taken seriously in nuclear physics.  The former
argument is that nucleon Z-graphs should be viewed in terms of their
constituents (quarks), and simultaneously reversing all three quark lines (in
$N \rightarrow \bar{N}$) is highly improbable; therefore pairs shouldn't be
taken seriously in nuclear physics.  These arguments are both correct and
irrelevant.

Degrees of freedom in physics are a choice, not an obligation.  A poor choice
hurts rather than helps.  That a field theory of (composite) nucleons, pions,
..., can be established seems not to be in doubt[8,9].  This effective field
theory can (and should) accommodate the chiral symmetry embedded in QCD, and
indeed forms a surrogate for the underlying QCD.  By choosing to ``freeze out''
d.o.f.\ corresponding to heavy mesons, etc., with a mass $\rsim \Lambda \sim$ 1
GeV, the large-mass QCD scale, the complexities of the short-range physics
(which surely require an appeal to QCD to understand fully) are reduced to
complicated nonlocal structures (in the same way that freezing out the photon
in QED leads to the retarded interaction between charged particles).  The short
range implies that {\it at low energies}, a nucleon in the act of exchanging a
heavy meson doesn't propagate very far, and consequently the nonlocal
structures can be expanded in terms of an infinite (and hopefully convergent)
series of {\it local} structures involving (derivatives of) delta
functions[10].  Given that a sensible field theory is possible, at least for
energies $\lsim \Lambda$, the size of high-energy reactions producing $N
\bar{N}$ pairs is irrelevant (since the energy is obviously greater than
$\Lambda$). For the low energies appropriate to the nuclear domain, the $N
\bar{N}$ pairs will be virtual and their effect is unphysical and manifestly
{\it unmeasurable}.  An analogous situation exists in electromagnetic
interactions, where a change of gauge can modify or even eliminate the amount
of ``pairs'' contributing to a given process, such as Compton scattering.  In a
theory of nucleons exhibiting chiral symmetry, a field redefinition (i.e., a
change of variables) corresponding to a chiral rotation can mix pion and
nucleon fields and can transform a PS type of coupling to PV coupling, for
example[11].  This involves a massive change of the ``pair'' structure, from
very large in magnitude to very small, and shows the unphysical nature of
``pairs'' at low energy.  One can also easily freeze out the pairs (via a
Foldy-Wouthuysen transformation) if one desires[11]. The results of freezing
d.o.f.\ will always make the dynamics have a more complicated form. This is
obvious in the nuclear physics context from the Feshbach (P,Q) theory of
reactions[12], where such freezing is also performed and leads to operators
with a much more complicated structure.

Finally, the calculability of reaction dynamics in few-nucleon systems
currently requires a nonrelativistic approach, although this restriction is
loosening.  The successful Dirac approach to calculating heavy nuclei
emphasizes the very strong (short-range) scalar- and vector-meson fields[13].
This leads to modest central forces (because of cancellation) and large
spin-orbit forces. In any perturbative estimate, relativistic effects are large
and the use of the Schr\"{o}dinger equation very dubious.  In few-nucleon
systems, however, although a consensus and complete understanding do not yet
exist, all estimates of these effects are small[14], though not negligible. 
Typically, changing from $T_{NR} = p^2/2M$ to $T_R = \sqrt{p^2 + M^2} - M$
produces 5\% changes in $\langle T \rangle$. Some care should be exercised,
however. For example, the effect on the extreme tail of the momentum
distribution is very large, because the shape is changed. Recently, two
calculations of the interaction-boost corrections (a particular type of
relativistic correction corresponding to an interacting pair of nucleons whose
center-of-mass is not the nuclear center-of-mass)  in the triton produced very
similar results ($\sim$250-300 keV repulsion) using very different conceptual
and calculational frameworks[14,15].  In addition, triton binding differences
between the Bonn-B potential and most other potential models ($\sim$300 keV)
were shown recently[16] to be due to a particular relativistic treatment of the
tensor operator in the Bonn-B OPEP. Thus, there is every expectation that
relativistic effects in few-nucleon systems are small and tractable, although
the confirmatory calculations remain to be performed. Because OPEP dominates
the nonrelativistic potential energy, it is {\it a priori} reasonable to 
assume that it also dominates the relativistic corrections to the interaction
energy.

Unfortunately, ``exact'' calculations are not tractable in heavier nuclei. It is
therefore difficult to assess the importance of individual elements of their
dynamics (as it was in few-nucleon systems prior to a decade ago). 
Nevertheless, it is safe to assume that the way in which calculations are
organized can play a significant role in the size of relativistic effects in
nuclei.  The use of potentials seems to minimize or hide large relativistic
corrections (i.e., large cancellations take place).

Another aspect is the strong short-range repulsion usually associated with the
vector meson field.  In an exact nonperturbative treatment, this leads to a
barrier between two nucleons that must be penetrated in order to have small
interparticle separations. This tends to strongly suppress the contribution of
the short-range repulsion to the total energy. In other words, although the
potential is very strong, the hole in the wave function that it produces
overcompensates and produces a small net result (i.e., the hole always
``wins'').  This suppression will also tend to reduce the overall effect of
relativity at small separations.  The latter is known to produce contributions
of $\sim$ 100 keV additional triton attraction in some nuclear force models[2].
In any event, the short-range structures in the potential produce relatively
little {\bf net} effect, as we shall show.

Finally, we note that chiral perturbation theory ($\chi$PT) (the effective
Lagrangian approach) suggests that the short-range potentials should be
comparable to the OPEP contribution[17].  In fact the former is suppressed due
to the barrier, leading to overall OPEP dominance.  Moreover, $\chi$PT suggests
a weak three-nucleon (or many-nucleon) force, and leads to a suppression of
higher-order perturbative contributions to the force (i.e., from loops). This
suggests that a sufficient understanding of the nuclear force to accommodate
recent calculational successes in the few-nucleon systems may be possible, 
although a strategy to achieve this is not yet evident.

\begin{figure}[htb]
  \epsfig{file=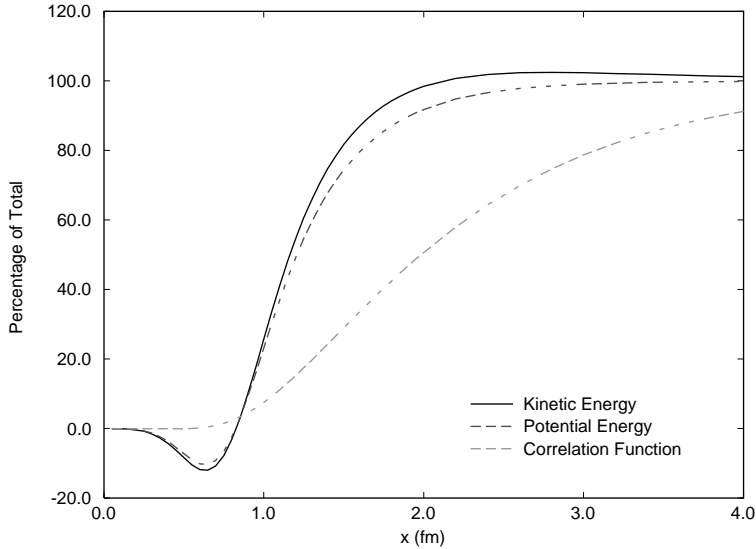,height=3.5in}
  \caption{Percentages of accrual of kinetic energy (solid line), potential  
energy (short dashed line), and probability (long dashed line) within an
interparticle separation, $x$, for any pair of nucleons.}
\end{figure}

Many of these facets of three-nucleon physics can be tied together using Figure~
(1).  One can ask the question, ``What fraction of the triton potential or
kinetic energy accrues within an interparticle separation, $x$, between any two
nucleons?''  This corresponds to forming
$$
\frac{\langle \Psi | \hat{O} \, \Theta (x - r_{12}) | \Psi \rangle}
{\langle \Psi | \hat{O}| \Psi \rangle}  \ ,
\eqno (1)
$$
where all three of the nucleon coordinates ${\bf r}_1, {\bf r}_2, {\bf r}_3$ 
(and hence ${\bf r}_{12}$) are integrated. For very large $x$, this quantity
approaches 100\%. Figure 1 shows the results for the original version of the
Argonne $V_{18}$ potential[2] (although other models that were examined are
virtually identical) for the kinetic energy $(\hat{O} = T)$, the potential
energy $(\hat{O} = V)$, and the correlation function $(\hat{O} = 1)$.  The
effect of the strong short-range repulsion is the dominant feature inside an
interparticle separation of 1 fm. Although the net potential energy (tracked by
the net kinetic energy) is repulsive inside 0.8 fm, the maximum accrual is
quite small, $\sim$~10\%.  The major accrual of attractive energy occurs
between 1.0 and 2.0 fm, which is the domain of OPEP and the TPEP tail.
A much cruder study of relativistic corrections in the alpha particle produces
a similar qualitative result[14].

This work demonstrates that some of the problems that we face are not as severe
as supposed and interprets a number of disparate results previously found.
Our conclusion is that OPEP dominance results from barrier impenetrability at
small NN separations.  This suppresses the short-range contributions relative
to what one might expect, reduces relativistic effects at short distances where
virtual momenta are very large, and suggests that relativistic corrections from
one-pion-exchange might be the dominant such physics. It further suggests that
an even better understanding of the origins of the observed triton binding
energy may be possible, allowing a credible separation of two-nucleon- and
three-nucleon-force mechanisms. We have contended that certain arguments
against the use of field theory to explicate the properties of the few-nucleon
systems are irrelevant to the nuclear domain, and that chiral-symmetry-based
field theory is the best hope for future progress in this area.

\begin{center}
{\bf ACKNOWLEDGEMENTS}\\
\end{center}

The work of JLF was performed under the auspices of the U.\ S.\ Department of
Energy, while that of GLP was supported in part by the U.\ S.\ Department of
Energy. One of us (JLF) would like to thank CEBAF and the Institute for Nuclear 
Theory for sponsoring a workshop on relativistic effects in nuclei, where many
of these ideas were addressed, and to H.\ Bethe for a stimulating discussion.

\pagebreak

\begin{center}
{\bf REFERENCES}
\end{center}

\begin{enumerate}

%1 
\item J.\ L.\ Friar, Summary talk presented at {\it XIV$^{\underline{th}}$ 
International Conference on Few-Body Problems in Physics,} Williamsburg, VA, 
May 31, 1994, ed.\ by F.\ Gross, ed.\ by F.\ Gross, AIP Conference Proceedings 
{\bf 334}, 323 (1995). This talk discusses the status of the field and possible 
future directions.

%2
\item J.\ L.\ Friar, G.\ L.\ Payne, V.\ G.\ J.\ Stoks, and J.\ J.\ de Swart,
Phys.\ Lett.\ {\bf B311}, 4 (1993).

%3
\item J.\ L.\ Friar, B.\ F.\ Gibson, and G.\ L.\ Payne, 
Phys.\ Rev.\ C {\bf 30}, 1084 (1984).

%4
\item R.\ B.\ Wiringa, Phys.\ Rev.\ C {\bf 43}, 1585 (1991).

%5
\item R.\ A.\ M.\ Klomp, V.\ G.\ J.\ Stoks, and J.\ J.\ de Swart,
Phys.\ Rev.\ C {\bf 44}, R1258 (1991); V.\ Stoks, R.\ Timmermans,
and J.\ J.\ de Swart, Phys.\ Rev.\ C {\bf 47}, 512 (1993).

%6
\item C.\ Ord\'o\~nez and U.\ van Kolck, Phys.\ Lett.\ {\bf B291}, 459 (1992); 
C.\ Ord\'o\~nez, L.\ Ray, and U.\ van Kolck, Phys.\ Rev.\ Lett.\ {\bf 72}, 
1982 (1994); U.\ van Kolck, Thesis, University of Texas, (1993).

%7
\item J.\ L.\ Friar and S.\ A.\ Coon, 
Phys.\ Rev.\ C {\bf 49}, 1272 (1994).

%8
\item S.\ Weinberg, Physica {\bf 96A}, 327 (1979); S. Weinberg, in
{\it Proceedings of the XXVI International Conference on High Energy Physics,
Volume I}, ed. by J.\ R.\ Sanford, AIP Conference Proceedings {\bf 272}, 346
(1993).

%9
\item J.\ Gasser and H.\ Leutwyler, Ann.\ Phys.\ (N.\ Y.\ )
{\bf 158}, 142 (1984); U.-G.\ Mei{\ss}ner, review talk at
{\it Third Workshop on High Energy Particle Physics},
Madras, India, January 1994, Preprint CRN--94/04. See the text
below equation (3) of the latter reference.

%10
\item G.\ P.\ Lepage, in {\it From Actions to Answers}, Proceedings of
the 1989 Theoretical Advanced Studies Institute in Elementary Particle Physics,
ed. by T.\ DeGrand and D. Toussaint, (World Scientific, Singapore, 1990), 
p. 483.

%11
\item  S.\ A.\ Coon and J.\ L.\ Friar, 
Phys.\ Rev.\ C {\bf 34}, 1060 (1986).

%12
\item F.\ S.\ Levin and H.\ Feshbach, {\it Reaction Dynamics},
(Gordon and Breach, New York, 1973).

%13
\item B.\ D.\ Serot, Repts.\ on Prog.\ in Physics {\bf 55}, 1855 (1992).

%14
\item J.\ Carlson, V.\ R.\ Pandharipande, and R.\ Schiavilla,
Phys.\ Rev.\ C {\bf 47}, 484 (1993).

%15
\item A.\ Stadler and F.\ Gross, Contributed paper presented at
{\it XIV$^{\underline{th}}$ International Conference on Few-Body Problems in
Physics}, Williamsburg, VA, May 31, 1994, ed.\ by F.\ Gross, p.\ 922.

%16
\item J.\ Adam, Invited talk presented at {\it XIV$^{\underline{th}}$
International Conference on Few-Body Problems in Physics}, Williamsburg, VA,
May 31, 1994, ed.\ by F.\ Gross, AIP Conference Proceedings (to appear); Y.\
Song and R.\ Machleidt, Contributed paper presented at {\it
XIV$^{\underline{th}}$ International Conference on Few-Body Problems in
Physics}, Williamsburg, VA, May 31, 1994, ed.\ by F.\ Gross, p.\ 189; A.\
Amghar and B.\ Desplanques, {\it ibid.}, p.\ 547.

%17
\item S.\ Weinberg, Nucl.\ Phys.\ {\bf B363}, 3 (1991); 
Phys.\ Lett.\ {\bf B251}, 288 (1990);
Phys.\ Lett.\ {\bf B295}, 114 (1992).

\end{enumerate}

\end{document}